\documentclass[reprint,amssymb,amsmath,superscriptaddress,aps,pra,nofootinbib]{revtex4-1}

\usepackage{color}
\usepackage{bm}
\usepackage{amsmath,amssymb}
\usepackage{amsfonts}
\usepackage{dsfont}
\usepackage{graphicx}
\usepackage{float}
\usepackage{subfigure}
\usepackage{verbatim}
\usepackage{amsfonts}
\usepackage{dcolumn}
\usepackage{bm}
\usepackage{color}
\usepackage[dvipsnames]{xcolor}
\usepackage{listings}
\definecolor{zaffre}{rgb}{0.0, 0.08, 0.66}
\usepackage[colorlinks=true,urlcolor=zaffre,citecolor=zaffre,linkcolor=zaffre]{hyperref}
\usepackage{mathrsfs}
\usepackage{filecontents}
\usepackage{mathtools}
\usepackage{times} 

\begin{document}

\title{Composable finite-size effects in free-space CV-QKD systems}

\author{Nedasadat Hosseinidehaj}  \email{n.hosseinidehaj@uq.edu.au}
\affiliation{Centre for Quantum Computation and Communication Technology, School of Mathematics and Physics, University of Queensland, St Lucia, Queensland 4072, Australia}

\author{Nathan Walk}
\affiliation{Dahlem Center for Complex Quantum Systems, Freie Universit{\"a}t Berlin, 14195 Berlin, Germany.}

\author{Timothy C. Ralph}
\affiliation{Centre for Quantum Computation and Communication Technology, School of Mathematics and Physics, University of Queensland, St Lucia, Queensland 4072, Australia}
\date{\today}

\begin{abstract}

Free-space channels provide the possibility of establishing continuous-variable quantum key distribution (CV-QKD) in global communication networks. However, the fluctuating nature of transmissivity in these channels introduces an extra noise which reduces the achievable secret key rate. 
We consider two classical post-processing strategies, post-selection of high-transmissivity data and data clusterization, to reduce the fluctuation-induced noise of the channel.  We undertake the first investigation of such strategies utilising a composable security proof in a realistic finite-size regime against both collective and individual attacks. We also present an efficient parameter estimation approach to estimate the effective Gaussian parameters over the post-selected data or the clustered data. Although the composable finite-size effects become more significant with the post-selection and clusterization both reducing the size of the data, our results show that these strategies are still able to enhance the finite-size key rate against both individual and collective attacks with a remarkable improvement against collective attacks--even moving the protocol from an insecure regime to a secure regime under certain conditions. 

\end{abstract}

\maketitle

\section{Introduction}
\label{sec1}

Quantum key distribution (QKD) \cite{Scarani.et.al.RVP.09,Xu.et.al.arxiv.19,Pirandola.et.al.arxiv.19} allows two trusted parties (traditionally called Alice and Bob) to share a secret key which is unknown to a potential eavesdropper (traditionally called Eve) by using quantum communication over an insecure quantum channel and classical communication over an authenticated classical channel. QKD systems were first proposed for discrete-variable quantum systems \cite{Bennett.Brassard.IEEE.84,Ekert.PRL.91}, where the key information is encoded onto the degrees of freedom of single photons, and the detection is realized by single-photon detectors, and then extended for continuous-variable (CV) quantum systems \cite{Ralph.PRA.99,Hillery.PRA.00,Reid.PRA.00}, where the key information is encoded onto the amplitude and phase quadratures of the quantized electromagnetic field of light, and detection is realised by (faster and more efficient) homodyne or heterodyne detectors. CV-QKD systems (see \cite{Garcia-Patron.PhD.07,Weedbrook.et.al.RVP.12,Diamanti.Leverrier.Entropy.15,Pirandola.et.al.arxiv.19} for review) have the potential of achieving higher secret key rates, as well as the advantage of compatibility with current telecommunication optical networks.

CV-QKD systems have experimentally been demonstrated over optical fibres \cite{experiment-CVQKD-2013,experiment-CVQKD-2016,chip-CVQKD-2019,commercial-fiber-CVQKD-2019,200km-CVQKD-2020}, however the maximum secure transmission distance is still limited to few hundred kilometres. As an alternative, free-space channels have the potential to extend the maximum transmission range of CV-QKD systems \cite{Hosseini.et.al.IEEE.19,Hosseini.Malaney.PRA1.15} as thay provide a possibility for the implementation of satellite-based QKD systems \cite{satellite-2017-1, satellite-2017-2}. However, in free-space channels (in contrast to optical fibers) the channel suffers from atmospheric turbulence (causing beam-wandering, beam shape deformation, beam broadening, etc.), which results in a random variation of channel transmissivity in time. This fluctuation effect can be characterized by a probability distribution of the channel transmissivity. Depending on the atmospheric effects, advanced probability distribution models have been proposed for the channel transmissivity \cite{Semenov.Vogel.PRA.09,Vasylyev.et.al.PRL.12,Vasylyev.et.al.PRL.16,Vasylyev.et.al.PRA.17,Vasylyev.et.al.PRA.18}, which accurately describe free-space experiments \cite{squeezing-freespace,Usenko-NJP-12}.

A free-space channel can be considered as a set of sub-channels, where the transmissivity of the channel is relatively stable for each sub-channel \cite{Usenko-NJP-12}. In a Gaussian CV-QKD protocol \cite{Garcia-Patron.PhD.07, Weedbrook.et.al.RVP.12}, Alice prepares Gaussian quantum states, which are modulated with a Gaussian distribution. Once these Gaussian states are transmited over a free-space channel to Bob, the fluctuating transmissivity of the channel makes the received state (at Bob's station) a non-Gaussian mixture of the Gaussian states obtained for each sub-channel \cite{Bohmann-Gaussianentanglement}. This non-Gaussian effect introduces an extra noise, which reduces the key rate \cite{Usenko-NJP-12,Hosseini.Malaney.ICC.15,NJP-2018}, however, the fluctuating transmissivity also provides a possibility to recover the key rate through the post-selection of data from sub-channels with high transmissivity \cite{Usenko-NJP-12}. The post-selection decreases the amount of channel fluctuation (i.e., decreases the variance of the transmissity distribution), which leads to a post-selected state with a more Gaussian nature (i.e., with less non-Gaussian noise). This post-selection has been shown to be effective for CV-QKD protocols in the asymptotic regime \cite{Usenko-NJP-12,Hosseini.Malaney.ICC.15} against Gaussian collective attacks\footnote{Note that the post-selection of transmission bins with high value has also been shown effective for enhancing the squeezing properties of light transmitted through the turbulent atmosphere \cite{squeezing-freespace,Vasylyev.et.al.PRL.16}, and improving the fidelity of the coherent-state teleportation over the turbulent atmosphere \cite{teleportation-freespace}.}. The other classical post-processing strategy which can also reduce the negative effect of the fluctuation-induced noise is to partition the recorded data into different clusters \cite{cluster-2019}, and analyse the security for each cluster separately. Although these strategies are both effective in reducing the fluctuation-induced noise, they also reduce the size of the effective data set used for the security analysis. Hence, since in practice a finite number of signals are exchanged between Alice and Bob, the composable finite-size issues become even more significant when either post-selection or clusterisation is applied. Thus, whether these classical post-processing strategies are still effective in a composable finite-size regime remains an open question.  

We consider the no-switching \cite{no-switching-2004, no-switching-2005} CV-QKD protocol (based on Gaussian-modulated coherent states and heterodyne detection) over a free-space channel. For the channel probability distribution we consider the elliptic-beam model \cite{Vasylyev.et.al.PRL.16}, which accounts for the deflection and deformation of a Gaussian beam caused by turbulence in atmospheric channels. We analyse the composable finite-size security of the protocol by using the recent security proof, stating that according to the Gaussian de Finetti reduction, for the no-switching protocol it is sufficient to consider Gaussian collective attacks in the finite-size, composable security proof \cite{Finite-size-Leverrier2017,Finite-size-Leverrier2019}. We answer the question whether the post-selection of high-tranmissivity sub-channels and data clusterisation can improve the performance of the free-space CV-QKD system in a composable finite-size regime against both collective and individual attacks\footnote{Note that when Eve has a restricted quantum memory, individual attacks can become the optimal eavesdropping attacks for the no-switching CV-QKD protocol \cite{Neda-Nathan-Tim}.}.

For both post-selection and clusterisation, the sub-channel transmissivity has to be estimated by publicly revealing a randomly-chosen subset of the data obtained over the stability time. There are other proposed methods \cite{channel-estimation-2012, channel-estimation-2015, channel-estimation-LO}, which utilize classical auxiliary probes or the local oscillator to estimate the sub-channel transmissivity. However, since these classical signals could be likely manipulated by Eve, the classical estimation of the channel may compromise the security. Note that in principle the parameters of the channel, i.e., transmissivity and excess noise, can be estimated for each sub-channel realization separately, however, in practice only a small number of signals can be transmitted over the stability time of the channel, which results in pessimistic error bars for the estimated parameters, which consequently overestimates Eve's information, and leads to a pessimistic bound for key rate. Hence, to estimate Eve's information, instead of estimating the parameters of each sub-channel, we utilize the data revealed over all sub-channels (which are used for the security analysis) to estimate the effective Gaussian parameters. 
Our security analysis shows that the optimised post-selection can improve the finite-size key rate against both individual and collective attacks. Our previous work on Gaussian post-selection \cite{postselection-Neda} showed a relatively modest improvement in the finite-size collective attacks in comparison with the significant improvement predicted by an asymptotic analysis. Surprisingly, our present work shows that the post-selection of high-transmissivity sub-channels provides a significant improvement in the composable finite-size regime, comparable to that predicted asymptotically.
Further, we show that the data clusterisation can also significantly improve the composable finite key rates against collective attacks, provided an optimal clusterisation of sub-channels are chosen. 

The structure of the remainder of the paper is as follows. In Sec. II, the no-switching CV-QKD system is described, with the security discussed in the composable finite-size regime.
In Sec. III, the CV-QKD system over free-space channels is discussed, with the security analysed using two approaches in the composable finite-size regime by introducing an efficient parameter estimation approach.  In Sec. IV, the finite-size composable security is analysed for the system with post-selection of high-transmissivity sub-channels and for the system with data clusterisation, and the significant improvement of the composable finite key rates against collective attacks using these strategies is illustrated. Finally, concluding remarks are provided in Sec. V.

\section{System model}\label{CV-QKD System}

We consider a Gaussian no-switching CV-QKD protocol \cite{no-switching-2004, no-switching-2005}, where Alice prepares Gaussian modulated coherent states and Bob uses heterodyne detection. In a prepare-and-measure scheme Alice generates two random real variables, $(a_q, a_p)$, drawn from two independent Gaussian distributions of variance $V_A$. Alice prepares coherent states by modulating a coherent laser source by amounts of $(a_q, a_p)$. The variance of the beam after the modulator is $V_A+1=V$ (where the 1 is for the shot noise variance), hence an average output state is thermal of variance $V$. The prepared coherent states are transmitted over an insecure quantum channel to Bob. For each incoming state, Bob uses heterodyne detection and measures both the $\hat q$ and $\hat p$ quadratures to obtain $(b_q, b_p)$. In this protocol, sifting is not needed, since both of the random variables generated by Alice are used for the key generation. When all the incoming quantum states have been measured by Bob, classical post-processing (including discretization, parameter estimation, error correction, and privacy amplification) over a public but authenticated classical channel is commenced to produce a shared secret key.

The prepare-and-measure scheme can be represented by an equivalent entanglement-based scheme \cite{Garcia-Patron.PhD.07,Weedbrook.et.al.RVP.12}, where Alice generates a pure two-mode squeezed vacuum state with the quadrature variance $V$.
Alice keeps one mode, while sending the second mode to Bob over the insecure quantum channel. When Alice applies a heterodyne detection to her mode, she projects the other mode onto a coherent state. At the output of the channel, Bob applies a heterodyne detection to the received mode.

\subsection{Composable Finite-size security analysis}


In the asymptotic regime collective attacks are as powerful as coherent attacks \cite{Renner.Cirac.PRL.2009}, and for Gaussian protocols, Gaussian collective attacks are asymptotically optimal \cite{Garcia-Patron.Cerf.PRL.06,Navascues.Grosshans.PRL.06,Wolf.Giedke.Cirac.PRL.06}. Note also that for Gaussian protocols, among individual attacks, Gaussian individual attacks are asymptotically optimal \cite{Garcia-Patron.PhD.07}.

In the finite-size regime, the no-switching CV-QKD protocol with $N$ coherent states sent by Alice to Bob is $\epsilon$-secure against Gaussian collective attacks in a reverse reconciliation scenario if $\epsilon {=} 2\epsilon_{\rm sm}{+}\bar \epsilon {+}\epsilon_{\rm PE}{+}\epsilon_{\rm cor}$ \cite{Finite-size-Leverrier-2015,Finite-size-Lupo-MDI} and if the key length $\ell$ is chosen such that \cite{Finite-size-Leverrier-2015,Finite-size-Lupo-MDI}
\begin{equation}\label{key-length-main}
\begin{array}{l}
 \ell  {\le} N'[ \beta I(a{:}b) {-} \chi^{\epsilon_{\rm PE}}(b{:}E) ] {-} \sqrt{N'} {\Delta _{\rm AEP}} {-} 2\log_2 (\frac{1}{{2\bar \epsilon }}),
 \end{array}
\end{equation}
where \cite{Finite-size-Leverrier-2015,Finite-size-Lupo-MDI}
\begin{equation}\label{delta-AEP}
\begin{array}{l}
\Delta_{\rm AEP} = (d{+}1)^2{+}4(d{+}1)\sqrt{\log_2({2{/}\epsilon_{\rm sm}^2})} {+}\\
\\ 2\log_2({2}{/}({\epsilon^2 \epsilon_{\rm sm}}))  {+} 4{\epsilon_{\rm sm}d}{/}{(\epsilon \sqrt{ N'}) },
 \end{array}
\end{equation}
where $N'=N-k$, with $k$ the number of data points Alice and Bob are required to disclose during the parameter estimation, $d$ is the discretization parameter (i.e., each symbol is encoded with $d$ bits of precision), $\epsilon_{\rm sm}$ is the smoothing parameter, $\epsilon_{\rm cor}$ and $\epsilon_{\rm PE}$ are the maximum failure probabilities for the error correction and parameter estimation, respectively, and $I(a{:}b)$ is the classical mutual information shared between Alice and Bob, and $0 \le \beta \le 1$ is the reconciliation efficiency. Note that in the finite-size regime the usual $\chi(b{:}E)$ (the maximum mutual information shared between Eve and Bob limited by the Holevo bound for the collective attack) has to be replaced by $\chi^{\epsilon_{\rm PE}}(b{:}E)$, taking into account the finite precision of the parameter estimation. In fact,  it is now assumed that Eve's information is upper bounded by $\chi^{\epsilon_{\rm PE}}(b{:}E)$, except with the probability $\epsilon_{\rm PE}$. The final key rate is then given by $\ell/N$.

Note that for the $\epsilon$-security analysis of the same protocol against Gaussian individual attacks we can still use Eq.~(\ref{key-length-main}), where $\chi^{\epsilon_{\rm PE}}(b{:}E)$ must be replaced by the classical mutual information between Eve and Bob, maximised by $I^{\epsilon_{\rm PE}}(b{:}E)$ except with the probability $\epsilon_{\rm PE}$.

Note also that based on the recent security proof in \cite{Finite-size-Leverrier2017,Finite-size-Leverrier2019}, for analysing the composable finite-size security of the no-switching CV-QKD protocol against general attacks, the security of the protocol can be first analysed against Gaussian collective attacks with a security parameter $\epsilon$ \cite{Finite-size-Leverrier-2015} through the use of Eq.~(\ref{key-length-main}), and then, by using the Gaussian de Finetti reduction \cite{Finite-size-Leverrier2017}, the security can be obtained against general attacks with a polynomially larger security parameter $\tilde \epsilon$ \cite{Finite-size-Leverrier2017}. Note that the security loss due to the reduction from general attacks to Gaussian collective attacks scales like $\mathit{O}(N^4)$ \cite{Finite-size-Leverrier2017}. More precisely, according to \cite{Finite-size-Leverrier2017}, $\epsilon$-security against Gaussian collective attacks implies $\tilde \epsilon$-security against general attacks, with $\tilde \epsilon / \epsilon = \mathit{O}(N^4)$.

\section{{Free-space CV-QKD systems}}

In free-space channels the atmospheric effects will cause the transmitted beam to experience fading. Hence, in contrast to a fiber link with a fixed transmissivity, the transmissivity, $\eta$, of a free-space channel fluctuates in time. Such fading channels can be characterized by a probability distribution $p(\eta)$ \cite{Usenko-NJP-12,Dong-PRA-10}. In fact, a fading channel can be decomposed into a set of sub-channels. Each sub-channel ${\eta_i}$ is defined as the set of events, for which the transmissivity is relatively stable, meaning that the fluctuations of the transmissivity is negligible. Each sub-channel ${\eta_i}$ occurs with probability $p_i$ so that $\sum\limits_{i} {p_i} = 1 $ or $\int_{0}^{\eta_{\rm max}} { p(\eta )} d\eta = 1$ for a continuous probability distribution, where $\eta_{\rm max}$ is the maximum realizable value of transmissivity of the fading channel. Thus, the Wigner function of the output state is the sum of the Wigner functions of the states after sub-channels weighted by sub-channel probabilities \cite{Dong-PRA-10}. Hence, the input Gaussian state $\rho_{\rm in}$ remains Gaussian after passing through each sub-channel, however, the resulting state at the output of the channel, $\rho_{\rm out} = \sum\limits_{i}{p_i \rho_i}$, (with $\rho_i$ the Gaussian state resulted from the transmission of the input Gaussian state $\rho_{\rm in}$ through the sub-channel $\eta_i$) is a non-Gaussian state \cite{Dong-PRA-10}.

In the equivalent entanglement-based scheme of the no-switching CV-QKD protocol, the initial pure two-mode Gaussian entangled state $\rho_{A{B_0}}$ with the quadrature variance $V$ is completely described by its first moment, which is zero, and its covariance matrix,
\begin{eqnarray}\label{AB0-CM}
{\bf{M}}_{A{B_0}}= \left[ {\begin{array}{*{20}{c}}
{V\,\bf{I}}&{\sqrt {{V^2} - 1} \,\bf{Z}}\\
{\sqrt {{V^2} - 1} \,\bf{Z}}&{V\,\bf{I}}
\end{array}} \right] .
\end{eqnarray}
Alice keeps mode~$A$ and sends mode ~${B_0}$ through an insecure free-space channel. After transmission of mode~${B_0}$ through a quantum sub-channel with transmissivity $\eta$ and excess noise $\xi_\eta$ (relative to the input of the sub-channel with transmissivity $\eta$), the covariance matrix of the \emph{Gaussian} state ${\rho_{{A{B_1}},\eta}}$ at the output of the sub-channel is given by
\begin{equation}\label{AB1-CM-subchannel}
{{\bf{M}}_{{A{B_1}},\eta}} = \left[ {\begin{array}{*{20}{c}}
{V\,\bf{I}}&{\sqrt {\eta \,} \sqrt {{V^2} - 1}\,\bf{Z}}\\
{\sqrt {\eta \,}  \,\sqrt {{V^2} - 1}\,\bf{Z}}&{\left[ {\eta (V - 1) + \eta \xi_\eta + 1  } \right]\,\bf{I}}
\end{array}} \right].
\end{equation}
Since the ensemble-average state at the output of a free-space channel, ${\rho_{{A{B_1}}}}$, is a non-Gaussian mixture of Gaussian states obtained from individual sub-channels, the elements of the covariance matrix of the ensemble-average state ${\rho_{{A{B_1}}}}$ are given by the convex sum of the moments given by Eq.~(\ref{AB1-CM-subchannel}). Hence, the covariance matrix of the \emph{non-Gaussian} ensemble-average state ${\rho_{A{B_1}}}$ at the output of the free-space channel is given by
\begin{equation}\label{AB1-CM-channel}
{{\bf{M}}_{A{B_1}}} {=} \left[ {\begin{array}{*{20}{c}}
{V\,\bf{I}}&{ \left\langle {\sqrt{\eta}} \right\rangle \sqrt {{V^2} - 1}\,\bf{Z}}\\
{ \left\langle {\sqrt{\eta}} \right\rangle  \,\sqrt {{V^2} - 1}\,\bf{Z}}&{\left[ { \left\langle {\eta} \right\rangle  (V - 1) + \left\langle {\eta \xi_\eta} \right\rangle + 1  } \right]\,\bf{I}}
\end{array}} \right].
\end{equation}
where the symbol $\left\langle .  \right\rangle $ denotes the mean value over the sub-channels (or over all possible values of $\eta$), i.e.,
\begin{equation}\label{mean}
\begin{array}{l}
{\left\langle \eta  \right\rangle } = \int_{0}^{\eta_{\rm max}} {\eta p(\eta )} d\eta ,\,\,{\left\langle {\sqrt \eta  } \right\rangle } = \int_{0}^{\eta_{\rm max}} {\sqrt \eta  p(\eta )} d\eta ,\, \\
\\
{\left\langle \eta \xi_\eta  \right\rangle } = \int_{0}^{\eta_{\rm max}} {\eta \xi_\eta p(\eta )} d\eta .
\end{array}
\end{equation}
Note that unlike the previous theoretical works on free-space CV-QKD \cite{Usenko-NJP-12,Hosseini.Malaney.ICC.15,NJP-2018,Hosseini.Malaney.PRA2.15, Hosseini.Malaney.QIC.17,Hosseini.Malaney.GLOBECOM.16, fast-fading} with the assumption of fixed excess noise, here we have assumed the channel excess noise can also randomly vary in time, where the value of the excess noise depends on the the value of the channel transmissivity.

From the covariance matrix of the non-Gaussian ensemble-average state in Eq.~(\ref{AB1-CM-channel}), it is evident that the fluctuating channel can be considered as a non-fluctuating channel with the effective transmissivity $\eta_f $ and effective excess noise $\xi_f$, so that the covariance matrix of the ensemble-average state can be rewritten as
\begin{equation}\label{effective}
\begin{array}{l}

{{\bf{M}}_{A{B_1}}} {=} \left[ {\begin{array}{*{20}{c}}
{V\,\bf{I}}&{ \sqrt{\eta_f} \sqrt {{V^2} - 1}\,\bf{Z}}\\
{ \sqrt{\eta_f}  \,\sqrt {{V^2} - 1}\,\bf{Z}}&{\left[
{  {\eta_f}( V {-} 1  ) {+} \eta_f \xi_f {+} 1
}

 \right]\,\bf{I}}
\end{array}} \right], {\rm{with}}
\\
\\
\eta_f = {\left\langle {\sqrt{\eta}} \right\rangle}^2,
\\
\\
\eta_f  \xi_f = \rm{Var}(\sqrt \eta  )(V - 1) + \langle {\eta \xi_\eta} \rangle, \,\,\,\,  \rm{and\,\,where} \\
\\
\rm{Var}(\sqrt \eta  ) = \left\langle \eta  \right\rangle  - {\left\langle {\sqrt \eta  } \right\rangle ^2}.
\end{array}
\end{equation}
According to Eq.~(\ref{effective}), the extra non-Gaussian noise, caused by the fluctuating nature of the channel, depends on the variance of the transmissivity fluctuations, $\rm{Var}(\sqrt \eta  )$, and the modulation variance, $V_A=V-1$.

\section{Composable finite-size Security analysis for free-space CV-QKD sytems}\label{Security}

Here we analyse the composable finite-size security of the no-switching CV-QKD protocol implemented over free-space channels using two approaches, first by analysing the security over all data, and second by analysing the security for each sub-channel separately. We also analyse the security against both general attacks (i.e., memory-assisted attacks) and individual attacks (i.e., non-memory attacks). 

\subsection{Security analysis over all data}

\subsubsection{General attacks}

Based on the leftover hash lemma \cite{lemma1,lemma2}, the number of approximately secure bits, $\ell$, that can be extracted from the raw key should be slightly smaller than the smooth min-entropy of Bob's string $b$ conditioned on Eve's system $E'$ (which characterizes Eve's quantum state $E$, as well as the public classical variable $C$ leaked during the QKD protocol), denoted by $H_{\min}^{\epsilon_{\rm sm}}(b|E')$ \cite{lemma1}, i.e., we have $\ell \le N' H_{\min}^{\epsilon_{\rm sm}}(b|E') {-} 2\log_2 (\frac{1}{{2\bar \epsilon }}) $, where $\bar \epsilon$ comes from the leftover hash lemma. Note that $N'$ indicates the length of Bob's string $b$ after the parameter estimation.  
The chain rule for the smooth min-entropy \cite{Finite-size-Leverrier-2015} gives $N'H_{\min}^{\epsilon_{\rm sm}}(b|E')=N'H_{\min}^{\epsilon_{\rm sm}}(b|EC) \ge N' H_{\min}^{\epsilon_{\rm sm}}(b|E) - \log_2|C|$, where $\log_2|C|=l_{\rm EC}$, with $l_{\rm EC}$ the size of data leakage during the error correction. Note that the leakage during the error correction can be given by $l_{\rm EC} = N' [H(b) - \beta I(a{:}b)]$ \cite{Finite-size-Leverrier-2015,Finite-size-Furrer,Finite-size-Lupo-MDI}, where $H(b)$ is Bob's Shannon entropy.
In order to calculate the length $\ell$ of the final key which is $\epsilon$-secure ($\epsilon {=} 2\epsilon_{\rm sm}{+}\bar \epsilon {+}\epsilon_{\rm PE}{+}\epsilon_{\rm cor}$ \cite{Finite-size-Leverrier-2015,Finite-size-Lupo-MDI}), the conditional smooth min-entropy $H_{\min}^{\epsilon_{\rm sm}}(b|E)$ has to be lower bounded when the protocol did not abort. 
Under the assumption of independent and identically distributed (i.i.d) attacks such as collective or individual attacks, where every signal transmitted is attacked with the same quantum operation, the asymptotic equipartition property \cite{Finite-size-Leverrier-2015, Marco-thesis, Marco} can be utilized to lower bound the conditional smooth min-entropy with the conditional von Neumann entropy. Explicitly, we have $N' H_{\min }^{\epsilon_{\rm sm}} (b\left| E \right.) \ge N' S(b\left| E \right.) - \sqrt{ N'}  \Delta_{\rm AEP}$ \cite{Finite-size-Leverrier-2015,Finite-size-Lupo-MDI}, where $S(b\left| E \right)$ is the conditional von Neumann entropy. The conditional von Neumann entropy $S(b\left| E \right)$ is given by $S(b|E)=H(b)- H^{\epsilon_{\rm PE}}(b{:}E)$, where Eve's information on Bob's string $b$ is upper bounded by $H^{\epsilon_{\rm PE}}(b{:}E)$, except with probability $\epsilon_{\rm PE}$ for a given attack (for collective attacks we have $H^{\epsilon_{\rm PE}}(b{:}E)=\chi^{\epsilon_{\rm PE}}(b{:}E)$ and for individual attacks we have $H^{\epsilon_{\rm PE}}(b{:}E)=I^{\epsilon_{\rm PE}}(b{:}E)$).

In our finite-size security analysis the assumption of collective attacks to lower bound the conditional smooth min-entropy comes with no loss of generality because based on the Gaussian de Finetti reduction, for the security analysis of the no-switching protocol against general attacks, it is sufficient to consider Gaussian collective attacks in the composable finite-size security proof \cite{Finite-size-Leverrier2017,Finite-size-Leverrier2019}.

Since the covariance matrix of the non-Gaussian ensemble-average state resulting from a free-space channel, given by Eq.~(\ref{effective}), can be described by the effective parameters $\eta_f$ and $\xi_f$, for the security analysis we can consider an optimal Gaussian collective attack with parameters $\eta_f$ and $\xi_f$. In this optimal attack Eve interacts individually with each transmitted signal through an optimal entangling cloner attack \cite{Entangling-cloner-2008} with the effective parameters $\eta_f$ and $\xi_f$, with her output ancillae stored in her quantum memory to be collectively measured later. Since this attack is i.i.d over all sub-channels, the conditional smooth min-entropy can be lower bounded by the conditional von Neumann entropy.  Thus the total finite-size key rate with the security parameter $\epsilon$ against Gaussian collective attacks is given by
\begin{equation}\label{key-rate-finite}
K^{\rm FS}_{\rm col} =  \frac{1}{N}\left[N'[ \beta I(a{:}b) {-} \chi^{\epsilon_{\rm PE}}(b{:}E) ] {-} \sqrt{N'} {\Delta _{\rm AEP}} {-} 2\log_2 (\frac{1}{{2\bar \epsilon }})\right].
\end{equation}
Note that we assume $N_s$ is the number of signals transmitted over each sub-channel, from which $k_s$ signals are revealed for the parameter estimation and $N'_s = N_s-k_s$ signals are used for the key generation. In total, a number of $N$ signal states are transmitted, from which $k$ signals are revealed over all sub-channels for the parameter estimation and $N' = N-k$ signals are used for the key generation. Note that in Eq.~(\ref{key-rate-finite}),  $I(a{:}b)$ is calculated based on the effective parameters $\eta_f $ and $\xi_f$, and Eve's information from collective attack, $\chi^{\epsilon_{\rm PE}}(b{:}E)$, is calculated based on the covariance matrix of the ensemble-average state, which can be estimated based on a relatively large number of signals $k$ (where $k  \gg  k_s$) revealed over all sub-channels (see Sec.~\ref{PE}). Note that according to \cite{Finite-size-Leverrier2017}, for the no-switching protocol, $\epsilon$-security against collective attacks implies $\tilde \epsilon$-security against general attacks, with $\tilde \epsilon / \epsilon = \mathit{O}(N'^4)$.


\subsubsection{Individual attacks}

Considering the fact that in reality Eve has access to a restricted quantum memory with limited coherence time, where each state stored into her quantum meory undergoes a specific amount of decoherence over the storage time, individual attacks might be more beneficial for Eve than collective attacks \cite{Neda-Nathan-Tim}. In terms of the interaction with the transmitted signals, an individual attack is the same as a collective attack, while in terms of the measurement Eve performs an individual measurement instead of a collective  measurement.  Among individual attacks, Gaussian attacks are also known to be optimal for the Gaussian CV-QKD protocols.

For the security analysis against Gaussian individual attacks we can also consider an optimal Gaussian individual attack with parameters $\eta_f$ and $\xi_f$. In this attack Eve interacts individually with each signal sent from Alice to Bob with the effective parameters $\eta_f$ and $\xi_f$ \footnote{Note that different schemes have been proposed for a Gaussian interaction in an optimal individual attack against the no-switching CV-QKD protocol, with the entangling cloner is one of them \cite{individula-2007-1,individula-2007-2}.}, with an individual measurement on her output ancillary state as soon as she obtains it. Note that in the no-switching CV-QKD protocol Eve does not need a quantum memory to perform the individual measurement, since there is no basis information withheld in this protocol. This individual attack is also i.i.d over all sub-channels, which means the conditional smooth min-entropy can be lower bounded by the conditional von Neumann entropy.  
Thus the total finite-size key rate with the security parameter $\epsilon$ against Gaussian individual attacks can also be given by Eq.~(\ref{key-rate-finite}), where $\chi^{\epsilon_{\rm PE}}(b{:}E)$ must be replaced by $I^{\epsilon_{\rm PE}}(b{:}E)$. Note that $I^{\epsilon_{\rm PE}}(b{:}E)$ has to be calculated based on the effective parameters of the channel, i.e., $\eta_f $ and $\xi_f$.

\subsection{Security analysis for each sub-channel separately}\label{subchannel-security}

For the security analysis against collective attacks with the security parameter $\epsilon$ ($\epsilon {=} 2\epsilon_{\rm sm}{+}\bar \epsilon {+}\epsilon_{\rm PE}{+}\epsilon_{\rm cor}$) one could also write $ H_{\min }^{\epsilon_{\rm sm}} (b\left| E \right) {=} \sum\nolimits_{i} p_i{H_{\min,i }^{\epsilon_{\rm sm},i}(b\left| E \right)} $, where $H_{\min,i}^{\epsilon_{\rm sm},i}(b\left| E \right)$ is the conditional smooth min-entropy for a sub-channel with parameters $\eta_i$ and $\xi_{\eta_i}$ occurring with probability $p_i$, and $\epsilon_{{\rm sm},i} = p_i \epsilon_{\rm sm}$. By considering an optimal Gaussian collective attack with parameters $\eta_i$ and $\xi_{\eta_i}$ over the sub-channel, one can lower bound $H_{\min,i}^{\epsilon_{\rm sm},i}(b\left| E \right)$ by the conditional von Neumann entropy since the attack is i.i.d over the sub-channel (note that this attack is not i.i.d over all sub-channels).  More explicitly, one can analyse the security for each sub-channel separately, i.e., calculate the composable finite-size key length for each sub-channel with the security parameter $\epsilon_i$ (where  $\epsilon_i = p_i \epsilon$) as $\ell_i =  N'_s[ \beta I_i(a{:}b) {-} \chi_i^{\epsilon_{\rm PE},i}(b{:}E) ] {-} \sqrt{N'_s} {\Delta _{\rm AEP}} {-} 2\log_2 (\frac{1}{{2\bar \epsilon_i }}) $ (where $N_s' = p_i N'$, and where $I_i(a{:}b)$ is the classical mutual information between Alice and Bob for the sub-channel, and  $\chi_i^{\epsilon_{\rm PE},i}(b{:}E)$ is Eve's information from collective attack over the sub-channel, which is calculated based on the covariance matrix ${{\bf{M}}_{A{B_1},\eta}}$ with parameter $\epsilon_{{\rm PE},i} = p_i \epsilon_{\rm PE}$), and then average over all sub-channels to obtain the total finite-size key rate with the security parameter $\epsilon$ as $\frac{1}{N} \sum\nolimits_{i} \ell_i $. Note that in a realistic finite-size regime this approach might result in pessimistic key rates. This is due to the fact that in practice only a small number of signal states can be transmitted over each sub-channel, which results in a very pessimistic finite key length for each sub-channel, i.e., $\ell_i$, since Eve's information, $\chi_i^{\epsilon_{\rm PE},i}(b{:}E)$, which is estimated based on a small number of signals $k_s$ (where $k_s = p_i k$), might be overestimated (when the block size is reduced, the error bar on the estimators of channel parameters increases, which results in estimating higher information for Eve). Note that this type of security analysis can also be used against individual attacks, however as discussed earlier the resulting key rate is expected to be pessimistic.

Note that in practice, it would be more practical to estimate the average SNR of the free-space channel based on the whole revealed data, and then choose an error-correction code rate based on this average SNR for the error correction of the whole remaining data. This means the mutual information should be calculated theoretically based on the effective parameters of the channel, i.e., $\eta_f $ and $\xi_f$. Alternatively and also ideally, it could be possible to estimate the SNR for each sub-channel separately, and then choose an error-correction code rate based on the sub-channel SNR for the error correction of the sub-channel data, which means the mutual information should be calculated theoretically by averaging over the mutual information obtained from each sub-channel as $\sum\nolimits_{i} p_i I_i (a{:}b) $.  However, estimation of SNR for each sub-channel based on a small number of signals revealed for each sub-channel does not give a good estimation of SNR. Note that in our numerical simulations we calculate the mutual information based on the effective parameters $\eta_f $ and $\xi_f$, which is a lower bound on $ \sum\nolimits_{i} p_i I_i (a{:}b)$. 

\subsection{Parameter estimation for free-space CV-QKD systems}\label{PE}

Alice and Bob are able to estimate the channel transmissivity and check its stability during the transmission of data \cite{cluster-2019, PE-freespace2019}. This is experimentally feasible, as the typical rate of free-space channel fluctuations is of the order of KHz, while the modulation and detection rate is typically of the order of several MHz, i.e., at least thousands of signal states can be transmitted during the stability time of the free-space channel \cite{Usenko-NJP-12}. The proper sub-channel estimation requires a large number of states to be sent through the channel during its stability. Then, some of the states for each sub-channel occurrence are randomly chosen for the parameter estimation.  

For instance, let us consider the free-space channel fluctuation rate of 1 KHz.  Then, we can assume that within each millisecond the channel is relatively stable and can be modelled with a fixed-transmissivity sub-channel of transmissivity $\eta$. Let us also consider the transmission and detection rate of 100 MHz. Hence, $N_s = 10^5$ signal states can be transmitted and detected at the receiver during the stability time of the channel. A fraction of these signals ($k_s = c N_s$) can be randomly chosen to reveal for parameter estimation, with the remaining data contributing to the secret key.
Finally, for instance, for 100 seconds of data transmission we will have transmitted $N = 10^{10}$ signal states (with $10^5$ signal states being transmitted during each stability time of the channel), with a fraction of which, $k = c N$, revealed over all sub-channels for the parameter estimation. Then, a number of $N'=N-k = (1-c) N$ signals will contribute to the shared secret key. Note that the security is not analysed for each sub-channel occurrence separately as it results in pessimistic key rates (see Sec.~\ref{subchannel-security}), instead the security is shown for the ensemble-average state, being obtained from the set of data of size $N - k $ upon all sub-channels.

Since in our security analysis it is sufficient to consider an optimal Gaussian attack with parameters $\eta_f$ and $\xi_f$, we can generalize the parameter estimation method introduced for a fixed-transmissivity quantum channel in \cite{MLE-estimator2010,MLE-estimator2012} to estimate the covariance matrix of the ensemble-average state using the data of size $k$ revealed over all sub-channels. For a no-switching CV-QKD protocol with Bob's heterodyne detector efficiency $\eta_B$ and electronic noise $\nu_B$, for a channel with fluctuating  transmissivity $\eta$, we can consider a normal linear model for Alice and Bob's correlated variables, $x_{A}$ and $x_{B}$, respectively.
\begin{equation}\label{eqp}
x_{B} = t  x_{A} + x_{n},
\end{equation}
where $t = \sqrt{\frac{\eta_B}{2}} {\langle { \sqrt{\eta}} \rangle} = \sqrt{\frac{\eta_B \eta_f}{2} }$, and $x_{n}$ follows a centred normal distribution with unknown variance $\sigma^2 = 1 + \nu_B +  \frac{\eta_B}{2} (\langle { \eta {\xi_\eta}} \rangle + {\rm{Var}}(\sqrt{\eta}) V_A) = 1 + \nu_B +  \frac{\eta_B}{2} \eta_f \xi_f$ (note that Alice's variable $x_{A}$ has the variance $V_A$). Using the total revealed data of size $k$, we can calculate the maximum-likelihood estimators for $  t  $ and $  \sigma^2 $, which are given by
\begin{equation}\label{b-eta}
\begin{array}{l}
\hat t = \frac{{\sum\nolimits_{i = 1}^{k} {{{A_{i}}}{{B_i}}} }}{{\sum\nolimits_{i = 1}^{k} {{{{{A_i}}}^2}} }}, \\
 \\
\hat  \sigma^2  = \frac{1}{{k}}\sum\nolimits_{i = 1}^{k} {{{({{B_{i}}} - \hat t {{A_{i}}})}^2}},
 \end{array}
\end{equation}
where $A_{i}$ and $B_{i}$ are the realizations of  $x_{A}$ and $x_{B}$, respectively.
The confidence intervals for these parameters are given by $t \in [\hat t-\Delta (t),\hat t+\Delta (t) ]$, and $ \sigma^2  \in [  \hat \sigma^2 -\Delta ( \sigma^2 ),\hat  \sigma^2 +\Delta ( \sigma^2 ) ]$ where
\begin{equation}\label{b-ci}
\begin{array}{l}
\Delta (t) = {z_{\epsilon_{\rm PE} /2}}\sqrt {\frac{{{  \hat \sigma ^2 }}}{{\sum\nolimits_{i = 1}^{k} {{{{{A}_{i}}}^2}} }}},  \\
\\
\Delta ({{\sigma }^2}) = {z_{\epsilon_{\rm PE} /2}}\frac{{{  \hat \sigma  ^2}\sqrt 2 }}{{\sqrt {k} }}.
 \end{array}
\end{equation}
Note that when no signal is exchanged, Bob's variable with realization $B_{0i}$ follows a centred normal distribution with unknown variance $\sigma_0^2 = 1 + \nu_B $, which is Bob's shot noise variance. The maximum-likelihood estimator for $\sigma_0^2$ is given by $\hat \sigma_0^2 = \frac{1}{{N}}\sum\nolimits_{i = 1}^{N} {{{{{B_{0i}}} }}}$. The confidence intervals for this parameters is given by $\sigma_0^2 \in [\hat \sigma_0^2-\Delta (\sigma_0^2),\hat \sigma_0^2+\Delta (\sigma_0^2) ]$, where $\Delta (\sigma_0^2)={z_{\epsilon_{\rm PE} /2}}\frac{{{\hat \sigma_0 ^2}\sqrt 2 }}{{\sqrt {N} }} $ \footnote{Note that ${z_{\epsilon_{\rm PE} /2}}$ is such that $1-{\rm erf}(\frac{{z_{\epsilon_{\rm PE} /2}}}{\sqrt{2}})/2 = \epsilon_{\rm PE} /2$.}.
Now we can estimate the effective parameters $\eta_f$  and $\xi_f $, which are given by
\begin{equation}\label{T-ci-average}
\begin{array}{l}
\hat \eta_f  = \frac{ 2  \hat t ^2 }{{{\hat \eta_B}}  },\\
\\
\Delta (\eta_f) =  \hat \eta_f \left( { \left| {\frac{{ 2{\Delta( t )}}}{{{\hat  t }}}} \right| + \left| {\frac{{\Delta (\eta_B )}}{{ \hat \eta_B }}} \right| } \right),
\\
\\
\hat \xi_f =  2\frac{\hat  \sigma^2  -\hat \sigma_0^2}{ \hat \eta_f \hat \eta_B},\\
\\
\Delta (\xi_f) = \hat \xi_f  {\left( {\left| {\frac{{\Delta ({ \sigma ^2 })}}{{{\hat \sigma^2 } - \hat \sigma _0^2}}} \right| + \left| {\frac{{\Delta (\sigma _0^2)}}{{{  \hat \sigma^2 } - \hat \sigma _0^2}}} \right| + \left| {\frac{{\Delta (\eta_B )}}{{\hat \eta_B }}} \right| + \left| {\frac{{\Delta (\eta_f )}}{{\hat \eta_f }}} \right|} \right)},
\end{array}
\end{equation}
where $\hat \eta_B$ is the estimator of Bob's detector efficiency with uncertainty $\Delta (\eta_B )$. 
Note that in order to maximise Eve's information from collective and individual attacks, the worst-case estimators of the effective parameters $\eta_f$  and $\xi_f $ should be used to evaluate Eve's information.



Note that for the sub-channel post-selection which we discuss in the next section, Alice and Bob also need to estimate the transmissivity of each sub-channel separately. We can use the similar method as discussed above to estimate the sub-channel transmissivity.
Considering a normal linear model $x_{B} = t_s  x_{A} + x_{n,s}$ for a sub-channel with transmissivity $\eta$, the maximum-likelihood estimators for the sub-channel parameters, $t_s$ and $\sigma^2_s$ (i.e., the variance of $x_{n,s}$), are given by \cite{MLE-estimator2010,MLE-estimator2012}
\begin{equation}\label{b-eta-subchannel}
\begin{array}{l}
 \hat t_s = \frac{{\sum\nolimits_{i = 1}^{k_s} {{{A_{i}}}{{B_i}}} }}{{\sum\nolimits_{i = 1}^{k_s} {{{{{A_i}}}^2}} }}, \\
 \\
 {\hat \sigma^2_s} = \frac{1}{{k_s}}\sum\nolimits_{i = 1}^{k_s} {{{({{B_{i}}} - \hat t_s {{A_{i}}})}^2}},
 \end{array}
\end{equation}
where $A_{i}$ and $B_{i}$ are the realizations of  $x_{A}$ and $x_{B}$ for the sub-channel, respectively, and $k_s $ is the number of signals revealed for the sub-channel.
The error bar for these parameters are given by
\begin{equation}\label{b-ci-subchannel}
\begin{array}{l}
\Delta (t_s) = {z_{\epsilon_{\rm PE} /2}}\sqrt {\frac{{{\hat \sigma ^2_s}}}{{\sum\nolimits_{i = 1}^{k_s} {{{{{A}_{i}}}^2}} }}},  \\
\\
\Delta ({{\sigma }^2_s}) = {z_{\epsilon_{\rm PE} /2}}\frac{{{\hat \sigma ^2_s}\sqrt 2 }}{{\sqrt {k_s} }}.
 \end{array}
\end{equation}
The worst-case estimator of the sub-channel transmissivity $\eta$ is then given by
\begin{equation}\label{T-ci-subchannel}
\begin{array}{l}
\eta^{\rm min} = \hat \eta - \Delta (\eta), {\rm where} \\
\\
\hat \eta = \frac{2  \hat t_s^2 }{{\hat \eta_B}  },\\
\\
\Delta (\eta) = \hat \eta\left( { \left| {\frac{{ {2\Delta(t_s)}}}{{{\hat t_s}}}} \right| + \left| {\frac{{\Delta (\eta_B )}}{{\hat \eta_B }}} \right| } \right),
 \end{array}
\end{equation}
Note that Alice and Bob have to perform their post-selection based on $\eta^{\rm min} = \hat \eta - \Delta (\eta)$.

\section{Classical post-processing strategies to improve free-space CV-QKD systems}

If we compare a fluctuating channel with an equivalent fixed-transmissivity channel with transmissivity $\eta_f = \langle \sqrt{\eta} \rangle^2$ and excess-noise $\langle {\eta \xi_{\eta}} \rangle$, the fluctuating channel has an extra non-Gaussian noise of ${\rm{Var}}(\sqrt{\eta}) (V-1)$ (see Eq.~(\ref{effective})), which reduces the key rate. Although fluctuating transmissivity of a free-space channel reduces the key rate, it also provides the possibility to improve or even recover it through the post-selection of sub-channels with high transmissivity \cite{Usenko-NJP-12} or the clusterisation of sub-channels \cite{cluster-2019}.

In the post-selection technique as introduced in \cite{Usenko-NJP-12}, the data collected for each sub-channel is kept, conditioned on the estimated sub-channel transmissivity being larger than a post-selection threshold $\eta_{\rm th}$, and discarded otherwise. In this technique, the security should be analysed over the post-selected data. With such a post-selection, the post-selected data becomes more Gaussian and more strongly correlated, since the post-selection reduces the fluctuation variance of the channel, while increases the average transmissivity of the channel. 

In the clusterisation technique as introduced in \cite{cluster-2019}, for the classical post-processing, Alice and Bob partition their data into $n$ different clusters, and perform classical post-processing (including reconciliation
and privacy amplification) over each cluster separately. The clusterization we consider here is such that the $j$th cluster $(j = 1, 2, ...,  n)$ corresponds to the $j$th channel transmissivity bin $(j - 1)\delta < \eta < j \delta$, with the bin size $\delta = \frac{\eta_{\rm max}}{n}$. Note that the clusterisation we consider here is the uniform binning of the probability distribution, however, in principle the width of each cluster can be optimised depending on the probability distribution. With such a technique, the clusterised data becomes more Gaussian, since the fluctuation variance of the channel is reduced within each cluster.

The post-selection has been shown to improve the free-space CV-QKD performance in terms of the key rate in the asymptotic regime against Gaussian collective attacks \cite{Usenko-NJP-12,Hosseini.Malaney.ICC.15}, and the clusterisation has been shown to improve the key rate in the finite-size regime against Gaussian collective attacks \cite{Usenko-NJP-12,Hosseini.Malaney.ICC.15}, but not in a composable-security regime.  However, both post-selection and  clusterisation reduces the size of the data used for the security analysis and the size of the data used for the parameter estimation. Hence, the composable finite-size effects become more significant in these scenarios. In the following sections we investigate the effectiveness of the post-selection and clusterisation in the composable finite-size regime against both individual and collective attacks (where the security against general attacks can be obtained by the security against collective attacks with a larger security parameter).

\subsection{Composable finite-size security analysis for the post-selection}

In the finite-size regime, the size of the post-selected data is $N_{\rm ps} = P_s N$, where $P_s$ is the post-selection success probability, i.e., the total probability for the channel transmissivity to fall within the post-selected region, $\eta \ge \eta_{\rm th}$, and is given by
$
{P_s} = \int_{{\eta _{\rm th}}}^{\eta_{\rm max}} {p(\eta )} d\eta
$.
Note that since in the post-selection protocol Eve's information should be estimated based on the post-selected data, Alice and Bob can only use the revealed data over the post-selected sub-channels to estimate the covariance matrix of the post-selected ensemble-average state, which means a data of size  $k_{\rm ps} = P_s k$ is used for the parameter estimation. Recall that $k$ is the amount of revealed data over all sub-channels.  Hence, the data of size $N'_{\rm ps}=N_{\rm ps} -k_{\rm ps} $ contributes to the post-selected key.  Explicitly, the finite-size key length of the post-selection protocol which is $\epsilon$-secure against Gaussian collective attacks in the reverse reconciliation scenario is given by
\begin{equation}\label{key-length-PS-collective}
\begin{array}{l}
 \ell^{\rm col}_{\rm ps}  \le N'_{\rm ps}[ \beta I_{\rm ps}(a{:}b){-}\chi^{\epsilon_{\rm PE}}_{\rm ps}(b{:}E) ] {-} \sqrt{ N'_{\rm ps}} {\Delta _{\rm AEP}} {-} 2\log_2 (\frac{1}{{2\bar \epsilon }}).
 \end{array}
\end{equation}
Eve's information from Gaussian collective attack in the post-selection protocol, is calculated based on the covariance matrix of the post-selected ensemble-average state ${\rho^{\rm ps}_{A{B_1}}}$, which is given by
\begin{equation}\label{effective-PS}
\begin{array}{l}

{{\bf{M}}^{\rm ps}_{A{B_1}}} {=} \left[ {\begin{array}{*{20}{c}}
{V\,\bf{I}}&{ \sqrt{\eta^{\rm ps}_f} \sqrt {{V^2} - 1}\,\bf{Z}}\\
{ \sqrt{\eta^{\rm ps}_f}  \,\sqrt {{V^2} - 1}\,\bf{Z}}&{\left[
{  {\eta^{\rm ps}_f}( V {-} 1  ) {+} \eta^{\rm ps}_f \xi^{\rm ps}_f {+} 1
}

 \right]\,\bf{I}}
\end{array}} \right],
\\
\\
\eta^{\rm ps}_{f} = {\left\langle {\sqrt{\eta}} \right\rangle}_{\rm ps}^2,
\\
\\
\eta^{\rm ps}_f \xi^{\rm ps}_f  = \rm{Var}_{\rm ps}(\sqrt \eta  )(V - 1) + \langle {\eta \xi_\eta} \rangle_{\rm ps}, \\
\\
\rm{Var}_{\rm ps}(\sqrt \eta  ) = \left\langle \eta  \right\rangle_{\rm ps}  - {\left\langle {\sqrt \eta  } \right\rangle _{\rm ps}^2}.
\end{array}
\end{equation}
where the symbol $\left\langle .  \right\rangle _{\rm ps}$ denotes the mean value over the post-selected sub-channels, i.e.,
\begin{equation}\label{mean-PS}
\begin{array}{l}
{\left\langle \eta  \right\rangle _{\rm ps}} = \frac{1}{P_s} \int_{{\eta _{\rm th}}}^{\eta_{\rm max}} {\eta p(\eta )} d\eta ,\,\,{\left\langle {\sqrt \eta  } \right\rangle _{\rm ps}} = \frac{1}{P_s}\int_{{\eta _{\rm th}}}^{\eta_{\rm max}} {\sqrt \eta  p(\eta )} d\eta ,\, \\
\\
{\left\langle \eta \xi_\eta  \right\rangle }_{\rm ps} = \frac{1}{P_s} \int_{\eta _{\rm th}}^{\eta_{\rm max}} {\eta \xi_\eta p(\eta )} d\eta .
 \end{array}
\end{equation}
Similarly, the finite-size key length of the post-selection protocol which is $\epsilon$-secure against Gaussian individual attacks in the reverse reconciliation scenario is given by
\begin{equation}\label{key-length-PS-individula}
\begin{array}{l}
 \ell^{\rm ind}_{\rm ps}  \le N'_{\rm ps}[ \beta I_{\rm ps}(a{:}b){-}I^{\epsilon_{\rm PE}}_{\rm ps}(b{:}E) ] {-} \sqrt{ N'_{\rm ps}} {\Delta _{\rm AEP}} {-} 2\log_2 (\frac{1}{{2\bar \epsilon }}),
 \end{array}
\end{equation}
where Eve's information from Gaussian individual attack in the post-selection protocol has to also be calculated based on the effective parameters $\eta^{\rm ps}_f $ and $\xi^{\rm ps}_f$.
Note that Eve's information $\chi^{\epsilon_{\rm PE}}_{\rm ps}(b{:}E)$ and $I^{\epsilon_{\rm PE}}_{\rm ps}(b{:}E)$ should be calculated based on the worst-case estimators of the effective parameters $\eta^{\rm ps}_f$  and $\xi^{\rm ps}_f $, where the estimations in Eqs.~(\ref{b-eta}) to (\ref{T-ci-average}) have to be calculated based on the revealed data over the post-selected sub-channels of size $k_{\rm ps}$. Note also that the classical mutual information between Alice and Bob obtained from the post-selection, $I_{\rm ps}(a{:}b)$, is calculated based on the effective parameters $\eta^{\rm ps}_f $ and $\xi^{\rm ps}_f$, and ${\Delta _{\rm AEP}}$ is calculated using Eq.~(\ref{delta-AEP}) with $N'$ being replaced by $N'_{\rm ps}$. Finally, the finite-size key rate of the post-selection protocol is given by $\ell_{\rm ps}/N$.
See Appendix~\ref{AppendixA} for the detailed calculation of Eve's information and Alice and Bob's mutual information.  

Note that the post-selection of high-transmissivity sub-channels has two different effects on the finite-size key rate. In fact, on the positive side the post-selection makes the ensemble-average state more Gaussian and more strongly correlated, while on the negative side the post-selection reduces the block size, and also makes the error bars larger in the parameter estimation.

As discussed earlier in Sec.~\ref{PE}, Alice and Bob can estimate the transmissivity of each sub-channel by revealing a fraction of the data allocated to each sub-channel. In the post-selection protocol, based on such an estimation of sub-channel transmissivity, they decide to keep or discard the sub-channel data. Note that the sub-channel transmissivity can be estimated using different schemes, e.g., by transmitting auxiliary coherent (classical) light probe signals that are intertwined with the quantum information \cite{channel-estimation-2012, channel-estimation-2015}, or by monitoring the local oscillator at the receiver, where the signal and the local oscillator have been sent in two orthogonally polarized modes through the free-space channel \cite{channel-estimation-LO}. However, in all these scenarios it would be very likely for Eve to manipulate the classical probe signal or the local oscillator in such a way to gain an advantage, for instance, forcing Alice and Bob to post-select a particular sub-channel, which is not actually within the post-selection region, can result in underestimating Eve's information by Alice and Bob.

\begin{figure*}[!t]
  \centering
  \subfigure{\includegraphics[width=7in]{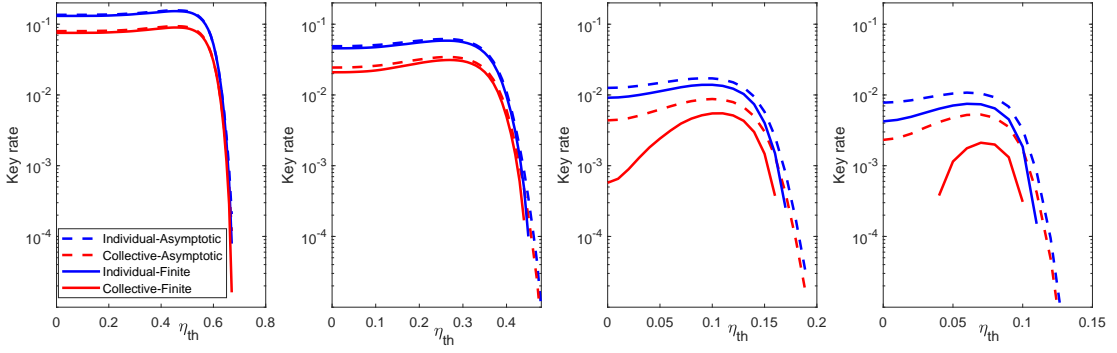}}
   \caption{Post-selected key rate in the asymptotic (dashed lines) and composable finite-size (solid lines) regime as a function of the post-selection threshhold $\eta_{\rm th}$, secure against collective (red lines) and individual (blue lines) attacks. The numerical values for the finite-size regime are the securiy parameter $\epsilon=10^{-9}$, with the parameter estimation being analysed for $\epsilon_{\rm PE}=10^{-10}$, and the discretization parameter $d=5$. The other parameters are chosen from the most recent CV-QKD experiment \cite{200km-CVQKD-2020} as follows. Bob's detector has efficieny $\eta_B = 0.6$, and electronic noise $\nu_{B} = 0.25$. The reconciliation efficiency is considered to be $\beta=0.98$. The expected excess noise of each sub-channel is assumed to be fixed as $\xi = 0.01$ \footnote{Note that although in our numerical simulations we have assumed a fixed excess noise for each sub-channel, our parameter estimation presented in Sec.~\ref{PE} also works for the case of fluctuating excess noise}. The block size is chosen to be $N=10^{10}$, half of which is used in total for parameter estimation. Since the non-Gaussian noise (i.e., the term $\rm{Var}(\sqrt{\eta}  )(V-1)$ in Eq.~(\ref{effective})) depends on the modulation variance, the modulation variance is optimized for each post-selection threshold to maximise the key rate secure against collective and individual attacks. We consider a probability  distribution for the free-space channel given by the elliptic-beam model (see Appendix~\ref{AppendixB} for more details on the model). From left to right we have the average $\langle { \eta  } \rangle  = 0.54, 0.32, 0.12, 0.08$, $\langle { \sqrt{\eta}  } \rangle  = 0.73, 0.56, 0.34, 0.27$, the variance $\rm{Var}(\sqrt{\eta}  ) = 0.003, 0.005, 0.003, 0.002$, and the maximum transmissivity $\eta_{\rm max} = 0.68, 0.46, 0.20, 0.13$ (see Fig.~{\ref{PDT} in Appendix for the corresponding probability distributions $p(\eta)$}). }\label{PS}
\end{figure*}

Fig.~\ref{PS} shows the post-selected key rate in both the asymptotic and composable finite-size regime as a function of the post-selection threshold $\eta_{\rm th}$, where the security is analysed against both collective and individual attacks. As can be seen for both asymptotic and finite-size regime, the key rate resulting from both collective and individual attacks first improves up to an optimized value, as the threshold value increases, and then the key rate decreases. As the threshold value increases, the variance of the channel fluctuations, $\rm{Var}(\sqrt{\eta}  )$ decreases, while the effective efficiency, $\langle { \sqrt \eta  } \rangle$ increases. As a result, the post-selected state becomes more Gaussian (i.e., with less non-Gaussian noise)  and more strongly correlated, which increases the mutual information between Alice and Bob. However, this increase in the mutual information happens at the cost of lower success probability, $P_s$, and larger error bars for the estimated parameters. Hence, there is an optimal threshold value which maximizes the post-selected key rate. As can be seen, from left to right, the post-selection becomes more effective. In fact, this type of post-selection is more useful for recovering the key rate in cases where it was strongly diminished by the free-space channel. Fig.~\ref{PS} also shows the significant improvement of the finite-size key rate from collective attacks due to the post-selection compared to the asymptotic regime. While without the post-selection, positive finite key rates cannot be generated against collective attacks for $\langle { \eta  } \rangle  = 0.08$, by performing the post-selection beyond $\eta_{\rm th}=0.05$, Alice and Bob are able to move from an insecure regime to a secure regime, and generate non-trivial positive finite key rates.

\subsection{Composable finite-size security analysis for the clusterisation}

\begin{figure*}[!t]
  \centering
  \subfigure{\includegraphics[width=7in]{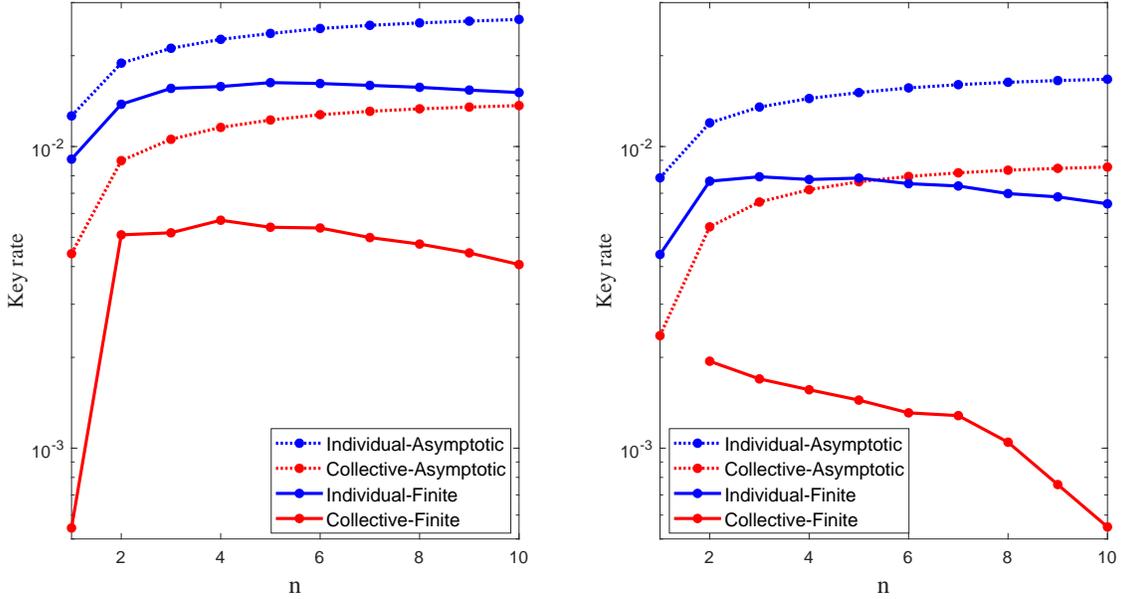}}
   \caption{Key rate in the asymptotic (dotted lines) and composable finite-size (solid lines) regime as a function of the number of clusters $n$, secure against collective (red lines) and individual (blue lines) attacks. The numerical values are the same as Fig.~\ref{PS}. From left to right we have the average $\langle { \eta  } \rangle  = 0.12, 0.08$, $\langle { \sqrt{\eta}  } \rangle  = 0.34, 0.27$, the variance $\rm{Var}(\sqrt{\eta}  ) = 0.003, 0.002$, and the maximum transmissivity $\eta_{\rm max} = 0.20, 0.13$. }\label{cluster}
\end{figure*}

For the clusterisation technique, in order to compute the key rate with security parameter $\epsilon$ (where $\epsilon = 2 \epsilon_{\rm sm} + \bar \epsilon + \epsilon_{\rm PE} + \epsilon_{\rm cor}$), the conditional smooth min-entropy $H_{\min }^{\epsilon_{\rm sm}} (b| E)$ can be written as the convex sum of the conditional smooth min-entropy of $n$ different clusters of data, i.e., $H_{\min }^{\epsilon_{\rm sm}} (b| E) = \sum_{j=1}^{n} {P_j \,\, H_{\min,j }^{\epsilon_{\rm sm},j} (b| E)} $, with $P_j =  \int_{\frac{(j-1) \eta_{\rm max}}{n}}^{\frac{j  \eta_{\rm max}}{n}} { p(\eta )} d\eta$ is the probability for the channel transmissivity to fall within the $j$th cluster, and $\epsilon_{{\rm sm},j} = P_j \epsilon_{\rm sm}$. Note that for each cluster Eve's attack can be considered as an i.i.d Gaussian attack with effective parameters $\eta^{j}_{f}$ and $\xi^{ j}_f$, given by 
\begin{equation}\label{effective-cluster}
\begin{array}{l}

\eta^{j}_{f} = {\left\langle {\sqrt{\eta}} \right\rangle}_{j}^2,
\\
\\
\eta^{j}_f \xi^{j}_f  = {\rm{Var}}_{j}(\sqrt \eta  )(V - 1) + \langle {\eta \xi_\eta} \rangle_{j}, \\
\\
{\rm{Var}}_{j}(\sqrt \eta  ) = \left\langle \eta  \right\rangle_{j}  - {\left\langle {\sqrt \eta  } \right\rangle _{j}^2},
\end{array}
\end{equation}
where the symbol $\left\langle .  \right\rangle _{j}$ denotes the mean value over all sub-channels within the $j$th cluster, i.e.,  
\begin{equation}\label{mean-cluster}
\begin{array}{l}
{\left\langle \eta  \right\rangle _{j}} = \frac{1}{P_j} \int_{\frac{(j-1) \eta_{\rm max}}{n}}^{\frac{j  \eta_{\rm max}}{n}} {\eta p(\eta )} d\eta ,
\\
\\
{\left\langle {\sqrt \eta  } \right\rangle _{j}} = \frac{1}{P_j}\int_{\frac{(j-1) \eta_{\rm max}}{n}}^{\frac{j  \eta_{\rm max}}{n}} {\sqrt \eta  p(\eta )} d\eta, \\
\\
\\
{\left\langle \eta \xi_\eta  \right\rangle }_{j} = \frac{1}{P_j} \int_{\frac{(j-1) \eta_{\rm max}}{n}}^{\frac{j  \eta_{\rm max}}{n}} {\eta \xi_\eta p(\eta )} d\eta .
 \end{array}
\end{equation}
Since the attack can be considered i.i.d over each cluster, we can lower bound $H_{\min,j}^{\epsilon_{\rm sm},j} (b| E) $ with the conditional von Neumann entropy and compute the key length with security parameter $\epsilon_j = P_j \epsilon$ for the $j$th cluster as $\ell^{\rm col}_j =  P_j N'[ \beta I_j(a{:}b) {-} \chi_j^{\epsilon_{\rm PE},j}(b{:}E) ] {-} \sqrt{P_j N'} {\Delta _{\rm AEP}} {-} 2\log_2 (\frac{1}{{2\bar \epsilon_j }}) $ against collective attacks, and $\ell^{\rm ind}_j =  P_j N'[ \beta I_j(a{:}b) {-} I_j^{\epsilon_{\rm PE},j}(b{:}E) ] {-} \sqrt{P_j N'} {\Delta _{\rm AEP}} {-} 2\log_2 (\frac{1}{{2\bar \epsilon_j }}) $ against individual attacks, where $I_j(a{:}b)$ is the classical mutual information between Alice and Bob for the $j$th cluster calculated based on the effective parameters $\eta^{j}_{f}$ and $\xi^{ j}_f$, and  $\chi_j^{\epsilon_{\rm PE},j}(b{:}E)$ is Eve's information from collective attack over the $j$th cluster, which is calculated based on the covariance matrix ${{\bf{M}}^{j}_{A{B_1}}}$
\begin{equation}\label{CM-cluster}
\begin{array}{l}

{{\bf{M}}^{j}_{A{B_1}}} {=} \left[ {\begin{array}{*{20}{c}}
{V\,\bf{I}}&{ \sqrt{\eta^{j}_f} \sqrt {{V^2} - 1}\,\bf{Z}}\\
{ \sqrt{\eta^{j}_f}  \,\sqrt {{V^2} - 1}\,\bf{Z}}&{\left[
{  {\eta^{j}_f}( V {-} 1  ) {+} \eta^{j}_f \xi^{j}_f {+} 1
}

 \right]\,\bf{I}}
\end{array}} \right],
\end{array}
\end{equation}
and  $I_j^{\epsilon_{\rm PE},j}(b{:}E)$ is Eve's information from individual attack over the $j$th cluster, which is calculated based on the effective parameters $\eta^{j}_{f}$ and $\xi^{ j}_f$. Note that Eve's information, $\chi_j^{\epsilon_{\rm PE},j}(b{:}E)$ and $I_j^{\epsilon_{\rm PE},j}(b{:}E)$, is now estimated based on the worst-case estimators of the effective parameters $\eta^{j}_f$  and $\xi^{j}_f $, where the estimations in Eqs.~(\ref{b-eta}) to (\ref{T-ci-average}) have to be calculated based on the revealed data over cluster $j$ of size $P_j k$, with the maximum failure probability $\epsilon_{{\rm PE},j} = P_j \epsilon_{\rm PE}$. Note also that $\Delta_{\rm AEP}$ is calculated using Eq.~(\ref{delta-AEP}) with $N'$ being replaced by $P_j N'$,  and $\Delta_{\rm AEP}$ is now calculated based on the parameters $\epsilon_j$, $\epsilon_{{\rm sm},j}$, and $\bar \epsilon_j = P_j \epsilon$. The total key rate with security parameter $\epsilon$ is then given by $\frac{1}{N} \sum_{j=1}^{n} \ell_j$.

Fig.~\ref{cluster} shows the key rate in both the asymptotic and composable finite-size regime secure against collective/individual attacks as a function of the number of clusters $n$. Note that $n=1$ indicates no clusterization, where security is analysed over all data. As can be seen clusterization always increases the asymptotic key rate against collective/individual attacks. However, by considering composable finite-size effects in the security analysis, there is an optimal number of clusters which maximises the key rate. In fact, as the number of clusters increases, the variance of channel fluctuation within each cluster decreases. As a result, the non-Gaussian noise becomes smaller for each cluster, which makes the state obtained over each cluster more Gaussian. However, as the number of clusters increases, the number of signals for each cluster decreases. As a result, the composable finite-size effects (i.e., the effect of $\Delta$-term, and the effect of parameter estimation which is now performed based on $P_j k$ signals with $\epsilon_{{\rm PE},j}$) become more significant, which reduces the key rate.       
Fig.~\ref{cluster} shows that while for $\langle { \eta  } \rangle  = 0.08$, without clusterization (i.e., $n=1$) the protocol is not secure against collective attacks in the composable finite-size regime, if Alice and Bob perform clusterisation as described above, the protocol becomes secure against collective attacks and finite key rate is maximised for $n=2$. 
Note that when the number of clusters, $n$, becomes sufficiently large, the security analysis is performed as if the security is analysed over each sub-channel separately (as described in Sec.~\ref{subchannel-security}). 

\section{Conclusions}

We have analysed the security of the no-switching CV-QKD protocol over free-space channels with fluctuating transmissivity in the composable finite-size regime against both collective and individual attacks. We introduced a parameter estimation approach, where Alice and Bob can efficiently estimate the effective parameters (i.e., the effective transmissivity and excess noise) of an optimal Gaussian attack using the data revealed over all sub-channels used for the security analysis. We analysed two classical post-processing strategies, the post-selection of high-transmissivity sub-channels and partitioning sub-channels into different clusters, in the composable finite-size regime, showing that these strategies can improve the finite-size key rate against both individual and collective attacks. Most remarkable improvement is for the finite-size collective attacks, which are the most practically relevant, where we see these classical post-processing allows significant key rates in situations that would otherwise be completely insecure.


\section{Acknowledgements}
The authors gratefully acknowledge valuable discussions with Andrew Lance, and Thomas Symul. This research is supported by the Australian Research Council (ARC) under
the Centre of Excellence for Quantum Computation and Communication Technology (Project No. CE170100012). NW acknowledges funding
support from the European Unions Horizon 2020 research
and innovation programme under the Marie Sklodowska-Curie grant agreement No.750905 and Q.Link.X from the
BMBF in Germany.

\appendix

\section{Key rate calculation}\label{AppendixA}

\subsection{Eve's information from collective attack}\label{chi-bE}

At the output of the channel Bob applies heterodyne detection to mode~$B_1$. Bob's heterodyne detector with efficiency $\eta_B$ and electronic noise variance of $\nu_B$ can be modeled by placing a beam splitter of transmissivity $\eta_B$ before an ideal heterodyne detector \cite{inefficient_homodyne, inefficient_heterodyne}. The heterodyne detector's electronic noise can be modelled by a two-mode squeezed vacuum state, $\rho_{{F_0}G}$, of quadrature variance $\upsilon $, where $\upsilon  = 1 + {2\nu_B}/(1 - \eta_B )$. One input port of the beam splitter is the received mode $B_1$, and the second input port is fed by one half of the entangled state $\rho_{{F_0}G}$, mode~$F_0$, while the output ports are mode~$B_{2}$ (which is measured by the ideal heterodyne detector) and mode $F$.

In a collective attack, Eve's information, $\chi(b{:}E)$, is given by $\chi(b{:}E)=\mathcal{S}(\rho_E)-\mathcal{S}(\rho_{E|B})$, where $\mathcal{S}(\rho )$ is the von Neumann entropy of the state $\rho$. Here we assume Bob's detection noise is not accessible to Eve. In this case $\mathcal{S}(\rho_E) = \mathcal{S}(\rho_{AB_1})$, where the entropy $\mathcal{S}(\rho_{AB_1})$ can be calculated through the symplectic eigenvalues $\nu _{1,2}$ of covariance matrix ${\bf{M}}_{A{B_1}}$\footnote{The von Neumann entropy of an $n$-mode Gaussian state $\rho$ with the covariance matrix $\bf{M}$ is given by $\mathcal{S}(\rho)=\sum\nolimits_{i = 1}^n {G(\frac{\nu _i-1}{2})} $, where $\nu _{i}$ are the symplectic eigenvalues of the covariance matrix ${\bf{M}}$, and $G(x) = (x + 1){\log _2}(x + 1) - x {\log _2}(x)$.} in Eq.~(\ref{effective}). The second entropy we require in order to determine $\chi(b{:}E)$ can be written as $\mathcal{S}(\rho_{E|B}) = \mathcal{S}(\rho_{E|B_2}) = \mathcal{S}(\rho_{AFG|B_2})$. The covariance matrix of the conditional state $\rho_{AFG|B_2}$ is given by ${\bf{M}}_{AFG|B_2}={\bf{M}}_{AFG}-{\boldsymbol{\sigma}}_{AFG,{B_2}} \,\, {\bf{H}}_{\rm het} \,\, \boldsymbol{\sigma}^T_{AFG,{B_2}}$, where ${\bf{H}}_{\rm het} = ({\bf{M}}_{B_2}+\bf{I})^{-1}$, and where ${\bf{M}}_{B_2} = {V_{B2}\bf{I}}$, where
\begin{equation}\label{V_B2}
V_{B2} = \eta_B [\eta_f(V-1)+ \eta_f \xi_f + 1] + (1-\eta_B)\upsilon.
\end{equation}
Note that the matrices ${\bf{M}}_{AFG}, \boldsymbol{\sigma}_{AFG,{B_2}}$, and ${\bf{M}}_{B_2}$ can be derived from the decomposition of the covariance matrix
\begin{equation}\label{big-CM}
{{\bf{M}}_{AFG{B_2}}} = \left[ {\begin{array}{*{20}{c}}
   {{{\bf{M}}_{AFG}}} & {\boldsymbol{\sigma} _{AFG,{B_2}}}  \\
   {{\boldsymbol{\sigma}^T_{AFG,{B_2}}}} & {{{\bf{M}}_{{B_2}}}}  \\
\end{array}} \right].
\end{equation}
Note that the covariance matrix ${\bf{M}}_{AFG{B_2}}$ is given by ${\bf{M}}_{AFG{B_2}} = ({\mathds{1}_A \oplus {\bf{S}}_{\rm bs} \oplus \mathds{1}_G})^T [{\bf{M}}_{A{B_1}}  \oplus {\bf{M}}_{{F_0}G} ]({\mathds{1}_A \oplus {\bf{S}}_{\rm bs} \oplus \mathds{1}_G})$, where ${\bf{S}}_{\rm bs}$ is the matrix for the beam splitter transformation (applied on modes $B_1$ and $F_0$), given by
\begin{eqnarray}\label{BS-CM}
{\bf{S}}_{\rm bs}= \left[ {\begin{array}{*{20}{c}}
{\sqrt{\eta_B}\,\bf{I}}&{\sqrt {1 - \eta_B} \,\bf{I}}\\
{-\sqrt {1 - \eta_B} \,\bf{I}}&{\sqrt{\eta_B}\,\bf{I}}
\end{array}} \right] ,
\end{eqnarray}
and the covariance matrix of the entangled state $\rho_{{F_0}G}$ is given by
\begin{eqnarray}\label{F0G-CM}
{\bf{M}}_{{F_0}G}= \left[ {\begin{array}{*{20}{c}}
{\upsilon\,\bf{I}}&{\sqrt {{\upsilon^2} - 1} \,\bf{Z}}\\
{\sqrt {{\upsilon^2} - 1} \,\bf{Z}}&{\upsilon\,\bf{I}}
\end{array}} \right] .
\end{eqnarray}
Note that in the finite-size regime, $\chi^{\epsilon_{\rm PE}}(b{:}E)$ should be calculated based on the worst-case estimators of $\eta_f$ and $\xi_f$. Note also that for the post-selection protocol, the parameters $\eta_f$ and $\xi_f$ have to be replaced by the post-selection parameters $\eta^{\rm ps}_f$ and $\xi^{\rm ps}_f$ from Eq.~(\ref{effective-PS}), and for the clusterisation, the parameters $\eta_f$ and $\xi_f$ have to be replaced by the cluster parameters $\eta^{j}_f$ and $\xi^{j}_f$ from Eq.~(\ref{effective-cluster}).

\subsection{Eve's information from individual attack}\label{I-bE}

Considering a free-space channel with effective parameters $\eta_f$ and $\xi_f$, defined in Eq.~(\ref{effective}), in the individual attack, Eve's information, $I(b{:}E)$, is given by $I(b{:}E) = {\log _2}\frac{{{V_{B_2^{\rm het}}}}}{{{V_{{B_2}^{\rm het}\left| E \right.}}}}$ \cite{individula-2007-1, individula-2007-2}, where $V_{B_2^{\rm het}}$ is the variance of heterodyne-detected mode $B_2$ for the post-selected sub-channel, and is given by $V_{B_2^{\rm het}}= (V_{B2}+1)/2$, where $V_{B2}$ is given in Eq.~(\ref{V_B2}). Note that $V_{{B_2^{\rm het}}|E}$ in the case of Bob's detection noise not being accessible to Eve is given by $V_{{B_2}^{\rm het}|E}=\eta_B[\frac{V{x_E}+1}{V+x_E}+\chi_{\rm het}]/2$, where $x_E=\eta_f(2-\xi_f)^2/(\sqrt{2-2\eta_f+\eta_f\xi_f}+\sqrt{\xi_f}))^2+1$, and $\chi_{\rm het} = [1+(1 - \eta_B ) + {2\nu_B}]/\eta_B $. Note that in the finite-size regime, $I^{\epsilon_{\rm PE}}(b{:}E)$ should be calculated based on the worst-case estimators of $\eta_f$ and $\xi_f$. Note also that the parameters $\eta_f$ and $\xi_f$ have to be replaced by parameters $\eta^{\rm ps}_f$ and $\xi^{\rm ps}_f$ from Eq.~(\ref{effective-PS}) for the post-selection protocol, and by parameters $\eta^{j}_f$ and $\xi^{j}_f$ from Eq.~(\ref{effective-cluster}) for cluster $j$.

\subsection{Mutual information between Alice and Bob}\label{I-ab}

The classical mutual information between Alice and Bob is given by $I(a{:}b) = {\log _2}\frac{{{V_{B_2^{\rm het}}}}}{{{V_{{B_2}^{\rm het}\left| A^{\rm het} \right.}}}}$.  The conditional variance $V_{{B_2}^{\rm het}|A^{\rm het}}$ is the variance of heterodyne-detected mode $B_2$ conditioned on Alice's heterodyne detection of mode $A$, which is given by $ V_{{B_2}^{\rm het}|A^{\rm het}} {=}\eta_B \eta_f (1+\chi_{\rm tot})/2$, where $\chi_{\rm tot}  = {\chi _{\rm line}} + \frac{{{\chi _{\rm het}}}}{\eta_f }$, with ${\chi _{\rm line}} = \xi_f - 1  + \frac{1}{\eta_f} $. Note that for the post-selected protocol, the parameters $\eta_f$ and $\xi_f$ have to be replaced by the post-selection parameters $\eta^{\rm ps}_f$ and $\xi^{\rm ps}_f$, and for the clusterisation, the parameters $\eta_f$ and $\xi_f$ have to be replaced by the cluster parameters $\eta^{j}_f$ and $\xi^{j}_f$.

\section{Elliptic-beam model}\label{AppendixB}

We have considered a free-space channel, where the probability distribution for the channel transmissivity is given by the elliptic-beam model \cite{Vasylyev.et.al.PRL.16}. This model can be used for an atmospheric channel including beam wandering, beam broadening and beam shape deformation \cite{Vasylyev.et.al.PRL.16}. However, for this model, referred to as the elliptic beam approximation, there is not an explicit form for the probability distribution. Here, we briefly discuss how to apply the model of the elliptic-beam approximation for calculation of the key rate. Further details on the model can be found in \cite{Vasylyev.et.al.PRL.16}. Within this model, it is assumed that turbulent disturbances along the propagation path result in beam wandering and deformation of the Gaussian beam profile into an elliptical form. The elliptic beam at the aperture plane is characterized by the beam-centroid position ${\bf{r}}_0 = (x_{0},y_{0})^T = (r_0 \cos \psi_0 , r_0 \sin \psi_0)^T$, and $W_1$ and $W_2$ as semi-axes of the elliptic spot, where the semi-axis $W_1$ has an angle $\psi \in [0,\pi/2)$ relative to the $x$ axis. Defining $\phi = \psi - \psi_0$, the aperture transmissivity $\eta_a$ is a function of real parameters $[x_{0},y_{0},\Theta_{1}, \Theta_{2}, \phi]$ \cite{Vasylyev.et.al.PRA.17}, which are randomly changed by the atmosphere. Note that $\Theta_1$ and $\Theta_2$ are related to the semi-axes, as $W_j^2 = W_0^2 \exp(\Theta_j)$ \cite{Vasylyev.et.al.PRL.16} for $j=1,2$ with $W_0$ the initial beam-spot radius. Note also that random fluctuations of the beam-centroid position ${\bf{r}}_0$, i.e. the parameters $x_{0}$ and $y_{0}$ cause the effect of beam wandering. For the parameters $[x_{0},y_{0},\Theta_{1}, \Theta_{2}]$, we can assume a four-dimensional Gaussian distribution, and for the parameter $\phi$, by assuming isotropic turbulence, we can assume a uniform distribution in the interval $[0,\pi/2]$ \cite{Vasylyev.et.al.PRA.17}. Under the assumption of isotropic turbulence, there is no correlation between $\phi$ with other linear parameters \cite{Vasylyev.et.al.PRL.16}. We also assume that $\left\langle {\bf{r}}_0 \right\rangle=0$, i.e., beam wandering fluctuations are placed around the reference-frame origin. Under this assumption, correlations between $x_0$, $y_0$ and $\Theta_j$ vanish \cite{Vasylyev.et.al.PRL.16}. Hence, we first generate $n$ independent Gaussian random vectors ${\bf{v}}_i = (x_{0i},y_{0i},\Theta_{1i}, \Theta_{2i}) $, $i=1,...,n$ and $n$ random uniformly-distributed angles $\phi_i \in [0,\pi/2]$. The Gaussian random parameters $(x_{0i},y_{0i},\Theta_{1i}, \Theta_{2i})$ can be characterized by the covariance matrix,
\begin{equation}\label{free-space-CM}
{\bf{M}} = \left( {\begin{array}{*{20}{c}}
{\left\langle {\Delta x_0^2} \right\rangle }&0&0&0\\
0&{\left\langle {\Delta y_0^2} \right\rangle }&0&0\\
0&0&{\left\langle {\Delta \Theta _1^2} \right\rangle }&{\left\langle {\Delta {\Theta _1}\Delta {\Theta _2}} \right\rangle }\\
0&0&{\left\langle {\Delta {\Theta _1}\Delta {\Theta _2}} \right\rangle }&{\left\langle {\Delta \Theta _2^2} \right\rangle }
\end{array}} \right)
\end{equation}
and the mean value $\left( {0,0,\left\langle {{\Theta _1}} \right\rangle ,\left\langle {{\Theta _2}} \right\rangle } \right)$. The elements of the covariance matrix and the mean values for weak turbulence are given by \cite{Vasylyev.et.al.PRL.16}
\begin{equation}\label{elements}
\begin{array}{l}
\left\langle {\Delta x_0^2} \right\rangle  = \left\langle {\Delta y_0^2} \right\rangle  = 0.33W_0^2\sigma _R^2{\Omega ^{ - \frac{7}{0}}},\\
\\
\left\langle {\Delta \Theta _1^2} \right\rangle  = \ln \left[ {1 + \frac{{1.2\sigma _R^2{\Omega ^{\frac{5}{6}}}}}{{{{\left( {1 + 2.96\sigma _R^2{\Omega ^{\frac{5}{6}}}} \right)}^2}}}} \right],\\
\\
\left\langle {\Delta {\Theta _1}\Delta {\Theta _2}} \right\rangle  = \ln \left[ {1 - \frac{{0.8\sigma _R^2{\Omega ^{\frac{5}{6}}}}}{{{{\left( {1 + 2.96\sigma _R^2{\Omega ^{\frac{5}{6}}}} \right)}^2}}}} \right],\\
\\
\left\langle {{\Theta _1}} \right\rangle  = \left\langle {{\Theta _2}} \right\rangle  = \ln \left[ {\frac{{{{\left( {1 + 2.96\sigma _R^2{\Omega ^{\frac{5}{6}}}} \right)}^2}}}{{{\Omega ^2}\sqrt {{{\left( {1 + 2.96\sigma _R^2{\Omega ^{\frac{5}{6}}}} \right)}^2} + 1.2\sigma _R^2{\Omega ^{\frac{5}{6}}}} }}} \right],
\end{array}
\end{equation}
where ${\sigma _R^2}$ is the Rytov parameter, $\Omega  = \frac{{kW_0^2}}{{2L}}$ is the Fresnel parameter, $k$ is the wave number and $L$ is the propagation distance.
After generating $n$ random vectors ${\bf{v}}_i = (x_{0i},y_{0i},\Theta_{1i}, \Theta_{2i}) $, and $n$ random angles $\phi_i \in [0,\pi/2)$, we can generate $n$ random transmissivity $\eta_{a,i}=\eta_a({\bf{v}}_i,\phi_i)$ as \cite{Vasylyev.et.al.PRL.16}
\begin{equation}\label{eta}
\eta_a({\bf{v}}_i,\phi_i)  = {\eta _0}\exp \left\{ { - {{\left[ {\frac{{{r_0}/a}}{{R\left( {\frac{2}{{{W_{\rm eff}(\phi)}}}} \right)}}} \right]}^{\lambda \left( {\frac{2}{{{W_{\rm eff}(\phi)}}}} \right)}}} \right\},
\end{equation}
where $r_0=\sqrt{x_0^2+y_0^2}$, is the distance between the beam and the aperture center, and $a$ is the radius of the circular receiver aperture. The transmissivity for the centered beam, i.e., for $r_0=0$ is given by \cite{Vasylyev.et.al.PRL.16}
\begin{equation}\label{eta_0}
\begin{array}{l}
{\eta _0} = 1 - {I_0}\left( {{a^2}\left[ {\frac{1}{{W_1^2}} - \frac{1}{{W_2^2}}} \right]} \right)\exp \left\{ { - {a^2}\left[ {\frac{1}{{W_1^2}} + \frac{1}{{W_2^2}}} \right]} \right\}\\
\\
 - 2\left[ {1 - \exp \left\{ { - \frac{{{a^2}}}{2}{{\left[ {\frac{1}{{{W_1}}} - \frac{1}{{{W_2}}}} \right]}^2}} \right\}} \right]\\
\\
 \times \exp \left\{ { - {{\left[ {\frac{{\frac{{{{\left( {{W_1} + {W_2}} \right)}^2}}}{{\left| {W_1^2 - W_2^2} \right|}}}}{{R\left( {\frac{1}{{{W_1}}} - \frac{1}{{{W_2}}}} \right)}}} \right]}^{\lambda \left( {\frac{1}{{{W_1}}} - \frac{1}{{{W_2}}}} \right)}}} \right\}.
\end{array}
\end{equation}
The further parameters, including effective squared spot radius, $W_{\rm eff}^2(\phi)$, the scale $R(\zeta )$  and shape $\lambda (\zeta )$ functions are given by are given by \cite{Vasylyev.et.al.PRL.16}
\begin{equation}\label{w_eff-R-lambda}
\begin{array}{l}
W_{\rm eff}^2(\phi) = 4{a^2} \times \\
\\
{\left[ {\mathcal{W}\left( {\frac{{4{a^2}}}{{{W_1}{W_2}}}\exp \left\{ {\frac{{{a^2}}}{{W_1^2}}(1 {+} 2{{\cos }^2}\phi )} \right\}\exp \left\{ {\frac{{{a^2}}}{{W_2^2}}(1 {+} 2{{\sin }^2}\phi )} \right\}} \right)} \right]^{ - 1}},\\
\\
R(\zeta ) = {\left( {\ln \left[ {2\frac{{1 - \exp \left\{ { - \frac{1}{2}{a^2}{\zeta ^2}} \right\}}}{{1 - \exp \left\{ { - {a^2}{\zeta ^2}} \right\}{I_0}({a^2}{\zeta ^2})}}} \right]} \right)^{ - \frac{1}{{\lambda (\zeta )}}}},\\
\\
\lambda (\zeta ) = 2{a^2}{\zeta ^2}\frac{{\exp \left\{ { - {a^2}{\zeta ^2}} \right\}{I_1}({a^2}{\zeta ^2})}}{{1 - \exp \left\{ { - {a^2}{\zeta ^2}} \right\}{I_0}({a^2}{\zeta ^2})}}\\
\\
 \times {\left( {\ln \left[ {2\frac{{1 - \exp \left\{ { - \frac{1}{2}{a^2}{\zeta ^2}} \right\}}}{{1 - \exp \left\{ { - {a^2}{\zeta ^2}} \right\}{I_0}({a^2}{\zeta ^2})}}} \right]} \right)^{ - 1}},
\end{array}
\end{equation}
where $\mathcal{W}(\zeta)$ is the Lambert $W$ function, and $I_j(\zeta)$ is the modified Bessel function of the $j$th order.

\begin{figure*}[!t]
  \centering
  \subfigure{\includegraphics[width=7in]{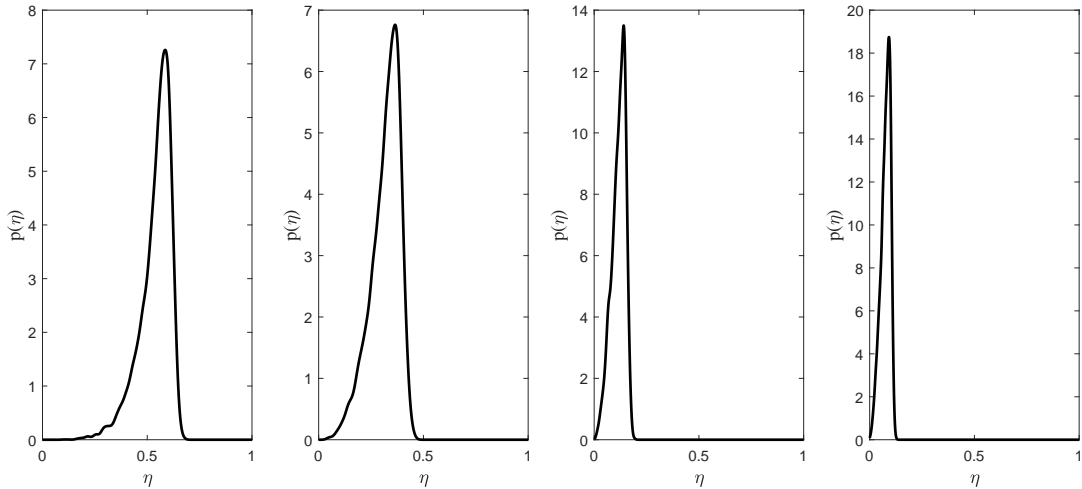}}
   \caption{The probability distribution function $p(\eta )$ obtained for the generated sampled data $\eta_i$, $n=10^4$. For the sample generations the parameters are chosen based on an experimentally implemented free-space experiment \cite{Usenko-NJP-12}. The elliptic-beam model \cite{Vasylyev.et.al.PRL.16} shows good agreement with the experimental distribution of the transmissivity \cite{Vasylyev.et.al.PRL.16}. The following parameter values are used, the wavelength $\lambda=809 $ nm, the initial beam-spot radius $W_0=20 $ mm,  deterministic attenuation $1.25 $ dB, and the radius of the receiver aperture $a = 40 $ mm. The Rytov parameter is given by ${\sigma _R^2}=1.23  C_n^2  k ^ {7/6}  L ^ {11/6}$, where we choose $C_n^2 = 1.5 \times 10^{-14} \, m^{-2/3} $, and the propagation distance $L = 1.5, 2, 3, 3.5$ km from left to right.}\label{PDT}
\end{figure*}

We also consider a deterministic (constant) tranmissivity $\eta_m \in [0, 1]$, which means the total transmissivity of the channel would be $\eta_i = \eta_m \eta_{a,i}$. Note that $\eta_m$ can be considered as the extinction factor of the atmospheric channel describing the absorption and scattering losses \cite{Vasylyev.et.al.PRA.17}. Based on the generated sampling data, one can estimate the mean value of any function of the transmissivity $f(\eta)$ as $\left\langle {f(\eta )} \right\rangle  = \frac{1}{n}\sum\limits_{i = 1}^n {f(\eta_i )}$. For instance, Eq.~(\ref{mean}) can be modified as
\begin{equation}\label{mean-MK}
\begin{array}{l}
{\left\langle \eta  \right\rangle } = \frac{1}{n}\sum\limits_{i = 1}^n {\eta_i } ,\,\,{\left\langle {\sqrt \eta  } \right\rangle } =  \frac{1}{n}\sum\limits_{i = 1}^n \sqrt{\eta_i } ,\, \\
\\
{\left\langle \eta \xi_\eta  \right\rangle } = \frac{1}{n}\sum\limits_{i = 1}^n {\eta_i \xi_{\eta_i} }.
\end{array}
\end{equation}
In our numerical analysis we have first fitted a probability distribution to the generated sampled data $\eta_i$ (shown in Fig.~\ref{PDT}), which gives us a numerical form for $p(\eta_i )$. Note that since no closed-form solution for $p(\eta )$ could be used, the integrals required to be computed for the security analysis (provided in Eqs.~(\ref{mean-PS}) and (\ref{mean-cluster})) should be numerically evaluated.

\end{document}